# Generalized continuum models for analysis of one-dimensional shear deformations in a structural interface with micro-rotations

Aleksey A. Vasiliev [a], Andrey E. Miroshnichenko [b,*], Sergey V. Dmitriev [c]

[a] *Department of Mathematical Modelling, Tver State University, 35 Sadoviy per., 170002 Tver, Russia*
[b] *Nonlinear Physics Centre, Research School of Physical Sciences and Engineering, The Australian National University, Canberra ACT 0200, Australia*
[c] *Institute for Metals Superplasticity Problems, Khalturin Str. 39, Ufa 450001, Russia*

**Abstract** Generalized continuum models for describing one-dimensional shear deformations of a Cosserat lattice are considered and their application to describing of structural effects essential for interfaces are discussed. The two-field long-wavelength micropolar model and its gradient and four-field generalizations are obtained and compared to the single-field conventional and gradient micropolar models. The single-field models can be applied to the analysis of long-wavelength deformations, but it does not describe short-wavelength waves and boundary effects. It is demonstrated that the two-field models describe both long-wavelength and short-wavelength harmonic waves and localized deformations and may be used in order to find stop band edges and to study the filtering properties of the interface. The two-field models make it possible to describe not only exponential but also short-wavelength boundary effects and evaluate degree of its spatial localization. The four-field model improves the two-field model in the description of the waves with wavenumbers in the middle part of the first Brillouin zone and may be useful to specify stop band edges in the case when minima/maxima of the dispersion curves belong to this region. The reported results are especially important for modeling of structural interfaces in the case when the length of localization is comparable with the interface thickness.

## 1. Introduction

Interfaces are often considered as surfaces of zero thickness. In some cases this assumption can be reasonable and may ease solution of physical problems. However, in reality interfaces possess a finite thickness and internal structure, which can play an important role for a number of problems and hence should be taken into account. The theory of structural interfaces of finite thickness has been developed by Bigoni and Movchan (2002), Bertoldi et al. (2007a, 2007b), and Brun et al. (2010). One of the attractive dynamical properties of structural interfaces used in various applications is their ability of elastic wave filtering. The boundary effects are usually not taken into consideration when stress-strain state of bodies of large sizes is analyzed. In the analysis of a thin interface one has to take into account the boundary effects because they typically affect regions that are comparable in size to the interface thickness. Particularly, boundary effects may be important in fracture problems.

One of the important problems of the theory of interfaces is the validity of different models for modeling structural effects. Classical models of continuum solid mechanics have a wide range of applications. Generalized continuum models have been developed for studying essential phenomena in structural solids that cannot be captured by the classical ones. For example, rotational degrees of freedom are incorporated in the micropolar theory (Cosserat and Cosserat, 1909; Eringen, 1968; Kim and Pizialy, 1987; Limat, 1988; Suiker et al., 2001; Pasternak and Muhlhaus, 2005; Pavlov et al., 2006; Potapov et al., 2009), higher-gradient effects are taken into account in the gradient theories (historical overview, discussion of models, recent studies and results can be found in Askes and Aifantis, 2011), the multi-field theory is developed in order to describe short-wavelength structural effects (see the review in Vasiliev et al., 2010b). For interfaces with a beam-like microstructure or when the finite size of particles is taken into account, not only translational displacements but also micro-rotations may be important for the analysis. Discrete one-dimensional model describing shear deformation of structural interface with micro-rotations has been considered in Vasiliev et al. (2010a). By combining approaches of the micropolar, gradient, and multi-field theories we derive the hierarchy of generalized continuum models for describing shear deformations of the structural interface.

This paper is organized as follows. In Section 2 we describe a discrete model of structural interface and derive a conventional long-wavelength micropolar model (Mm) and a gradient micropolar model (GMm). Two equivalent forms of long-wavelength two-field micropolar model (2FMm), a two-field gradient micropolar model (2FGMm) and a four-field micropolar model (4FMm) are obtained in Section 3. Harmonic and localized dynamical solutions are derived in Section 4. We compare accuracy of different continuum models; apply them to the problem of finding the stop band edges. Static exponential and short-wavelength boundary effects are analyzed in frame of different continuum models in Section 5. Section 6 concludes the paper.

## 2. Discrete and single-field micropolar models

### 2.1. Discrete model

We consider the structural interface of finite thickness presented by several layers of the Cosserat lattice of particles that possess not only translation displacements $\widetilde{u}_n$, $\widetilde{v}_n$, but also rotational degree of freedom $\widetilde{\varphi}_n$ (Fig. 1). Potential energy describing interaction of neighboring particles $k$ and $m$ is assumed in the form (Suiker et al., 2001)

$$E_{pot} = \frac{1}{2} K_n (\widetilde{u}_m - \widetilde{u}_k)^2 +$$

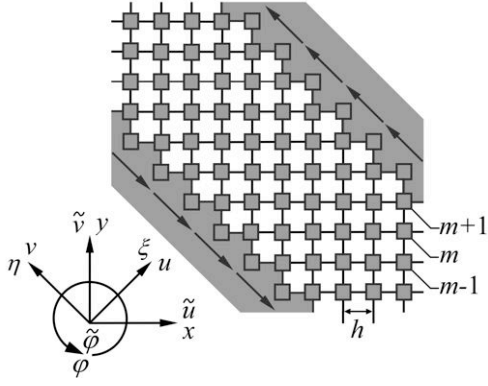

**Fig. 1.** Structural interface of finite thickness. Coordinate systems and notations.

$$+ \frac{1}{2} K_s \left( \tilde{v}_m - \tilde{v}_k - r_{k,m} \frac{\tilde{\varphi}_m + \tilde{\varphi}_k}{2} \right)^2 + \frac{1}{2} G_r (\tilde{\varphi}_m - \tilde{\varphi}_k)^2, \quad (1)$$

where $r_{k,m}$ is the length of connections; $K_n$ and $K_s$ are the stiffness coefficients of the bonds in longitudinal and transverse directions, respectively, and $G_r$ is the stiffness with respect to rotations.

Potential (1) is used in modeling of materials when finite size of particles is taken into account. It is widely used in models of granular media (Limat, 1988; Suiker et al., 2001; Pasternak and Muhlhaus, 2005). The potential of the form (1) as a discrete energy of the two disc interaction was used in Shufrin (2012) where the structures exhibiting negative Poisson's ratio are designed. It was used in development of a directed continuum model of micro- and nano-scale thin films in Randow et al. (2006). Similar model for particle interaction and microstructural parameters for some crystals were obtained from experimental data and reported in Pavlov et al. (2006), Potapov et al. (2009). Another type of materials where potential (1) is used are lattice structures (Ruzzene and Scarpa, 2003, 2005), materials with beam-like microstructure (Kim and Pizialy, 1987). Relations between the parameters $K_n$, $K_s$, and $G_r$ and microstructural parameters for complex connections of two particles and parameters usually used in beam models are given, for example, in Vasilive et al. (2010a).

Kinetic energy of the $n$-th particle is defined by the relation

$$E_{kin}^n = \frac{1}{2} M \dot{\tilde{u}}_n^2 + \frac{1}{2} M \dot{\tilde{v}}_n^2 + \frac{1}{2} J \dot{\tilde{\varphi}}_n^2,$$

where $M$ is the mass and $J$ is the moment of inertia of the particle.

Equations for components $u_m$ and $v_m$, $\varphi_m$ are decoupled in linear case of one-dimensional deformations. Modeling of longitudinal deformations of Cosserat lattice has been discussed in Vasiliev et al. (2008). In the present work we consider one-dimensional shear deformations (Fig. 1). Equations of motion for the components $v_m$, $\varphi_m$ in the coordinate system $O\xi\eta$ have the form

$$M\ddot{v}_m = K_n \Delta_{\xi\xi} v_m + K_s (\Delta_{\xi\xi} v_m - d\Delta_\xi \varphi_m),$$
$$J\ddot{\varphi}_m = (2G_r - K_s d^2)\Delta_{\xi\xi} \varphi_m + K_s d(\Delta_\xi v_m - 4d\varphi_m), \quad (2)$$

where $\Delta_\xi w_m = w_{m+1} - w_{m-1}$,

$\Delta_{\xi\xi} w_m = w_{m+1} - 2w_m + w_{m-1}$, and $d = \sqrt{2}h$ is the distance between rows of particles (Fig. 1).

*2.2. Micropolar model*

Vector function with components $V^{[1]}(\xi,t)$, $\Phi^{[1]}(\xi,t)$ is used in the single-field continuum model to describe the deformations of the discrete system. It is assumed that

$$V^{[1]}(\xi,t)\big|_{\xi=md} = v_m(t), \quad \Phi^{[1]}(\xi,t)\big|_{\xi=md} = \varphi_m(t).$$

Using these relations one can rewrite Eq. (2) in the functional-difference form for the field functions. Taylor series expansion of the displacements and rotations, taking into account up to the second order terms, results in the equations of Mm

$$MV^{[1]}_{tt} = K_n d^2 V^{[1]}_{\xi\xi} + d^2 K_s (V^{[1]}_{\xi\xi} - 2\Phi^{[1]}_\xi),$$
$$J\Phi^{[1]}_{tt} = (2G_r - K_s d^2) d^2 \Phi^{[1]}_{\xi\xi} + 2d^2 K_s (V^{[1]}_\xi - 2\Phi^{[1]}). \quad (3)$$

*2.3. Gradient micropolar model*

If up to the fourth order terms are taken into account in the Taylor series expansions of the displacements and rotations, we obtain the following equations of the GMm

$$MV^{[1]}_{tt} = (K_n + K_s)d^2 \left( V^{[1]}_{\xi\xi} + \frac{1}{12} d^2 V^{[1]}_{\xi\xi\xi\xi} \right)$$
$$- 2d^2 K_s \left( \Phi^{[1]}_\xi + \frac{1}{6} d^2 \Phi^{[1]}_{\xi\xi\xi} \right),$$
$$J\Phi^{[1]}_{tt} = (2G_r - K_s d^2) d^2 \left( \Phi^{[1]}_{\xi\xi} + \frac{1}{12} d^2 \Phi^{[1]}_{\xi\xi\xi\xi} \right)$$
$$+ 2d^2 K_s \left( V^{[1]}_\xi + \frac{1}{6} d^2 V^{[1]}_{\xi\xi\xi} - 2\Phi^{[1]} \right). \quad (4)$$

## 3. Multi-field micropolar models

*3.1. Two-field long-wavelength micropolar model*

Single-field models were constructed in Section 2 using single vector function with the same degrees of freedom that were chosen to describe kinematics of primitive cell. Considering a larger cell that includes several primitive lattice cells and using, correspondingly, a larger number of fields to describe the deformations of the body leads to the multi-field continuum models (Vasiliev and Miroshnichenko, 2005; Vasiliev et al., 2008; Gonella and Ruzzene, 2010; Vasiliev et al., 2010b). In order to derive equations of 2FMm, we consider a macrocell that includes two neighboring primitive cells. An additional index $s = 1, 2$ is introduced to mark displacements and rotations in odd and even cells, $v_m^{[s]}(t)$ and $\varphi_m^{[s]}(t)$. Defor-mations in this case are defined by four finite-difference equations. Two vector functions with components $v^{[s]}(\xi,t)$, $\varphi^{[s]}(\xi,t)$, $s=1,2$, are used to describe displacements and rotations of odd and even layers. By using Taylor series expansions of displacements and rotations up to the second order terms, we obtain the set of four coupled equations of the long-wavelength 2FMm

$$Mv_{tt}^{[s]} = K_n d^2 v_{\xi\xi}^{[s]} + K_s d^2 \left(v_{\xi\xi}^{[s]} - 2\varphi_\xi^{[s]}\right) +$$
$$+ (-1)^s \left[-K_n d^2 \Delta v_{\xi\xi} - K_s d^2 \left(\Delta v_{\xi\xi} - 2\Delta\varphi_\xi\right) - 2(K_n + K_s)\Delta v\right],$$
$$J\varphi_{tt}^{[s]} = \left(2G_r - K_s d^2\right) d^2 \varphi_{\xi\xi}^{[s]} + 2K_s d^2 \left(v_\xi^{[s]} - 2\varphi^{[s]}\right) +$$
$$+ (-1)^s \left[-\left(2G_r - K_s d^2\right) d^2 \Delta\varphi_{\xi\xi} - \right. \quad (5)$$
$$\left. - 2K_s d^2 \left(\Delta v_\xi - 2\Delta\varphi\right) - 2\left(2G_r + K_s d^2\right)\Delta\varphi\right]$$

where $s=1,2$ and notations $\Delta v = v^{[2]}(\xi,t) - v^{[1]}(\xi,t)$, $\Delta\varphi = \varphi^{[2]}(\xi,t) - \varphi^{[1]}(\xi,t)$ are used.

The first two terms in the right-hand side of (5) for the fields $v^{[s]}(\xi,t)$ and $\varphi^{[s]}(\xi,t)$ are the same as in the single-field Mm (3). The terms in square brackets describe coupling of the fields. The formulation of the model in the form (5) may be useful in the development of the phenomenological theories. One can formulate different models both for sublattices and for describing their interaction.

In terms of new field functions

$$V^{[1]} = \frac{1}{2}\left[v^{[2]} + v^{[1]}\right], \quad V^{[2]} = \frac{1}{2}\left[v^{[2]} - v^{[1]}\right],$$
$$\Phi^{[1]} = \frac{1}{2}\left[\varphi^{[2]} + \varphi^{[1]}\right], \quad \Phi^{[2]} = \frac{1}{2}\left[\varphi^{[2]} - \varphi^{[1]}\right], \quad (6)$$

the system (5) splits into two uncoupled systems. The set of equations for the fields $V^{[1]}(\xi,t)$ and $\Phi^{[1]}(\xi,t)$ coincides with the set of equations (3) of the single-field micropolar theory. The set of equations for $V^{[2]}(\xi,t)$ and $\Phi^{[2]}(\xi,t)$ has the form

$$MV_{tt}^{[2]} = -K_n d^2 V_{\xi\xi}^{[2]} - K_s d^2 \left(V_{\xi\xi}^{[2]} - 2\Phi_\xi^{[2]}\right) -$$
$$- 4(K_n + K_s)V^{[2]},$$
$$J\Phi_{tt}^{[2]} = -\left(2G_r - K_s d^2\right) d^2 \Phi_{\xi\xi}^{[2]} - \quad (7)$$
$$- 2K_s d^2 \left(V_\xi^{[2]} - 2\Phi^{[2]}\right) - 4\left(2G_r + K_s d^2\right)\Phi^{[2]}.$$

Presentation of 2FMm (5) in the form of two uncoupled systems (3) and (7) is more convenient for analysis. In particular, such a form explicitly demonstrates that the single-field Mm (3) is included into 2FMm.

As in the case of long-wavelength single-field model, we consider two approaches to improve the long-wavelength two-field model.

*3.2. Two-field gradient micropolar model*

If the fourth order derivatives are taken into account in the Taylor series expansions of the displacements and rotations in the equations for the macrocell that include two neighboring primitive cells, we come to the equations of 2FGMm. In this case, in addition to the equations of GMm (4), we have equations

$$MV_{tt}^{[2]} = -K_n d^2 \left(V_{\xi\xi}^{[2]} + \frac{1}{12} d^2 V_{\xi\xi\xi\xi}^{[2]}\right) -$$
$$- K_s d^2 \left[V_{\xi\xi}^{[2]} + \frac{1}{12} d^2 V_{\xi\xi\xi\xi}^{[2]} - 2\left(\Phi_\xi^{[2]} + \frac{1}{6} d^2 \Phi_{\xi\xi\xi}^{[2]}\right)\right] -$$
$$- 4(K_n + K_s)V^{[2]},$$
$$J\Phi_{tt}^{[2]} = -\left(2G_r - K_s d^2\right) d^2 \left(\Phi_{\xi\xi}^{[2]} + \frac{1}{12} d^2 \Phi_{\xi\xi\xi\xi}^{[2]}\right) -$$
$$- 2K_s d^2 \left(V_\xi^{[2]} + \frac{1}{6} d^2 V_{\xi\xi\xi}^{[2]} - 2\Phi^{[2]}\right) - \quad (8)$$
$$- 4\left(2G_r + K_s d^2\right)\Phi^{[2]}.$$

*3.3. Four-field micropolar model*

It is also possible to construct four-field model using four vector fields with components $v^{[s]}(\xi,t)$ and $\varphi^{[s]}(\xi,t)$, $s=\overline{1,4}$, to describe deformations of the lattice. Eight finite-difference equations define the displacements $v_m^{[s]}(t)$ and rotations $\varphi_m^{[s]}(t)$ of particles, $s=\overline{1,4}$, in the macrocell consisting of four primitive cells. Taylor series expansions up to the second order terms result in the eight coupled dynamic equations of long-wavelength 4FMm for the fields $v^{[s]}(\xi,t)$ and $\varphi^{[s]}(\xi,t)$, $s=\overline{1,4}$. In terms of new variables for displacements

$$V^{[1]} = \frac{1}{4}\left[v^{[2]} + v^{[1]} + v^{[4]} + v^{[3]}\right],$$
$$V^{[2]} = \frac{1}{4}\left[v^{[2]} - v^{[1]} + v^{[4]} - v^{[3]}\right],$$
$$V^{[3]} = \frac{1}{4}\left[v^{[2]} - v^{[1]} - \left(v^{[4]} - v^{[3]}\right)\right],$$
$$V^{[4]} = \frac{1}{4}\left[v^{[2]} + v^{[1]} - \left(v^{[4]} + v^{[3]}\right)\right],$$

and similar expressions for rotations, the set of eight equations splits into three independent subsets. Equations for the fields $V^{[1]}(\xi,t)$, $\Phi^{[1]}(\xi,t)$ and $V^{[2]}(\xi,t)$, $\Phi^{[2]}(\xi,t)$ are the equations (3) and (7) of 2FMm. In addition, we have the following four equations

$$MV_{tt}^{[3]} = -2K_n \left(dV_\xi^{[4]} + V^{[3]}\right) +$$
$$+ K_s \left(d^3 \Phi_{\xi\xi}^{[4]} - 2d\left[V_\xi^{[4]} - \Phi^{[4]}\right] - 2\Phi^{[3]}\right),$$
$$J\Phi_{tt}^{[3]} = dK_s \left(d^2 \left[2\Phi_\xi^{[4]} - V_{\xi\xi}^{[4]}\right] - 2d\Phi^{[3]} - 2V^{[4]}\right) -$$
$$- 4G_r \left(\Phi^{[3]} + d\Phi_\xi^{[4]}\right),$$
$$MV_{tt}^{[4]} = 2K_n \left(dV_\xi^{[3]} - V^{[4]}\right) -$$
$$- K_s \left(d^3 \Phi_{\xi\xi}^{[3]} - 2d\left[V_\xi^{[3]} - \Phi^{[3]}\right] + 2\Phi^{[4]}\right),$$

$$J\Phi_{tt}^{[4]} = dK_s\left(-d^2\left[2\Phi_\xi^{[3]} - V_{\xi\xi}^{[3]}\right] - 2d\Phi^{[4]} + 2V^{[3]}\right) - \\ - 4G_r\left(\Phi^{[4]} - d\Phi_\xi^{[3]}\right) \quad (9)$$

## 4. Dynamical solutions

### 4.1. Harmonic and localized dynamical solutions for lattice

The dynamical solutions to the discrete model (2) having the form

$$v_m(t) = e^{i\omega t + Km}\bar{v}, \quad \varphi_m(t) = e^{i\omega t + Km}\bar{\varphi} \quad (10)$$

have been analyzed in Vasiliev et al. (2010a). Harmonic solutions $K = iK_{Im}$ are usually analyzed for unbounded media (Kim and Pizialy, 1987; Suiker et al., 2001). We consider complex values of $K = K_{Re} + iK_{Im}$ with the possibility to analyze not only harmonic $(K_{Re} = 0)$ but also spatially localized $(K_{Re} \neq 0)$ solutions. The latter ones can be especially important for interfaces of relatively small thickness.

The substitution of (10) into equations (2) leads to the linear system

$$\begin{bmatrix} a_1 + M\omega^2 & -a_2 \\ a_2 & a_3 + J\omega^2 \end{bmatrix} \begin{bmatrix} \bar{v} \\ \bar{\varphi} \end{bmatrix} = \begin{bmatrix} 0 \\ 0 \end{bmatrix}, \quad (11)$$

where

$$a_1 = 2(K_n + K_s)(\cosh K - 1), \quad a_2 = 2K_s d \sinh K, \\ a_3 = 2(2G_r - K_s d^2)(\cosh K - 1) - 4K_s d^2. \quad (12)$$

The dispersion relations $\omega = \omega(K)$ can be found from the condition

$$\det\begin{bmatrix} a_1 + M\omega^2 & -a_2 \\ a_2 & a_3 + J\omega^2 \end{bmatrix} = 0. \quad (13)$$

Two branches of the harmonic wave dispersion curves exist in the plane $K_{Re} = 0$. They connect the points

$$\omega|_{K=0} = 0, \quad \omega|_{K=0} = \sqrt{4K_s d^2 / J}, \quad (14)$$

$$\omega|_{K=i\pi} = \sqrt{4(K_n + K_s)/M}, \quad \omega|_{K=i\pi} = \sqrt{8G_r/J} \quad (15)$$

on the lines $K_{Im} = 0$ and $K_{Im} = \pi$. The analysis shows that the considered discrete system supports both exponential and short-wavelength localized solutions with branches in the planes $K_{Im} = 0$ and $K_{Im} = \pi$.

### 4.2. Analysis of single-field models

Here we will consider the following continuum analogs of the discrete solutions (10)

$$V(\xi,t) = e^{i\omega t + K\xi/d}\bar{V}, \quad \Phi(\xi,t) = e^{i\omega t + K\xi/d}\bar{\Phi}. \quad (16)$$

Substituting (16) into equations (3) of Mm, we come to the system (11) and equation (13) for the dispersion curves with components

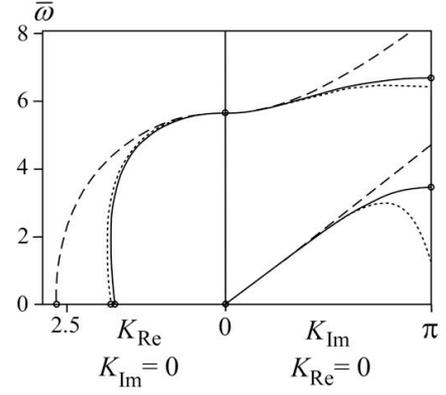

**Fig. 2.** Approximation of the discrete system dispersion curves (solid lines) by the dispersion curves of Mm and GMm (dashed and dotted lines, respectively) for harmonic $(K_{Re} = 0)$ and localized $(K_{Im} = 0)$ dynamical solutions. Approximations by the dispersion curves of 2Mm and 2GMm are given in Fig. 3a.

$$c_1 = (K_n + K_s)K^2, \quad c_2 = 2K_s dK, \\ c_3 = (2G_r - K_s d^2)K^2 - 4K_s d^2. \quad (17)$$

It can be easily checked that the components $c_m$ are the Taylor series expansions of the components $a_m$ [see (12)] of the discrete model up to the second order terms.

In the case of GMm, Eq. (4), components $c_m$ are the Taylor series expansions of $a_m$ up to the fourth order terms.

Taylor series expansions of the dispersion curves of the discrete and single-field continuum models at the point $(K_{Re}, K_{Im}) = (0, 0)$ in the planes $K_{Re} = 0$ and $K_{Im} = 0$ coincide up to terms of the second order in the case of Mm and up to terms of the fourth order for GMm. Hence, both models give exact values (14) and exact group velocities of waves. GMm also describes the dispersion of waves.

Figure 2 illustrates the results for various single-field models. Dispersion curves $\bar{\omega} = \omega\sqrt{M/K_s}$ of the discrete model, Mm and GMm in the planes $K_{Im} = 0$ and $K_{Re} = 0$ are depicted by the solid, dashed, and dotted lines, respectively. Here we use dimensionless parameters $\bar{J} = J/Md^2 = 1/8, \bar{K}_n = K_n/K_s = 2$, and $\bar{G}_r = G_r/K_s d^2 = 0.7$. Dispersion curves of single-field models are considered as approximations of the dispersion curves of the lattice in the region $0 \leq K_{Im} \leq \pi$ only (Kunin, 1982). All dispersion curves coincide at the points (14). The single-field models show a good accuracy for long-wavelength harmonic $(K_{Re} = 0)$ and exponentially localized $(K_{Im} = 0)$ solutions near the point $(K_{Im}, K_{Re}) = (0, 0)$. However, one can note that the single-field models give considerable error for short-wavelength harmonic solutions $(K_{Im} \approx \pi)$ and for the The multi-field approach efficiently solves the problem of modeling of short-wavelength waves.

2FMm includes Eq. (3) of Mm and thus, it can be used in the long-wavelength region $0 \leq K_{\text{Im}} < \pi/2$ with the same accuracy as Mm.

Substitution of expressions (16) into equations (7) of 2FMm leads to the system of equations (11) and equation (13) for the dispersion relations with components

$$\tilde{c}_1 = -(K_n + K_s)(K^2 + 4), \quad \tilde{c}_2 = -2K_s dK,$$
$$\tilde{c}_3 = -(2G_r - K_s d^2)(K^2 + 4) - 4K_s d^2. \quad (18)$$

These components can be derived by substitution $K \to -i\pi + K$ in $a_m$ followed by Taylor series expansions at the point $K = 0$ up to the second order terms. Taking into account this property of the dispersion relations, one can show that the dispersion curves of the two-field model defined in the region $0 \leq K_{\text{Im}} < \pi/2$, $K_{\text{Re}} \geq 0$, being reflected with respect to the plane $K_{\text{Im}} = \pi/2$ on the region $\pi/2 < K_{\text{Im}} \leq \pi$, $K_{\text{Re}} \geq 0$, approximate the dispersion curves of the discrete system in the area of short-wavelength waves at the point $(K_{\text{Im}}, K_{\text{Re}}) = (\pi, 0)$.

Thus, 2FMm provides a second-order approximation of the dispersion curves of harmonic and localized solutions in the planes $K_{\text{Re}} = 0$, $K_{\text{Im}} = 0$, and $K_{\text{Im}} = \pi$ at the points $(K_{\text{Im}}, K_{\text{Re}}) = (0,0)$ and $(K_{\text{Im}}, K_{\text{Re}}) = (\pi, 0)$.

2FGMm (4), (7) improves 2FMm. In this case the components $c_m$ are the Taylor series expansions of the components $a_m$ up to the fourth order terms, meaning that the dispersion curves are approximated with the fourth-order accuracy.

Figure 3 illustrates the results for the two-field models. Dispersion curves in the planes $K_{\text{Im}} = 0$, $K_{\text{Re}} = 0$, and values (15) on the line $K_{\text{Im}} = \pi$. Both single-field models do not reproduce branches of the dispersion curves corresponding to the localized short-wavelength solutions in the plane $K_{\text{Im}} = \pi$.

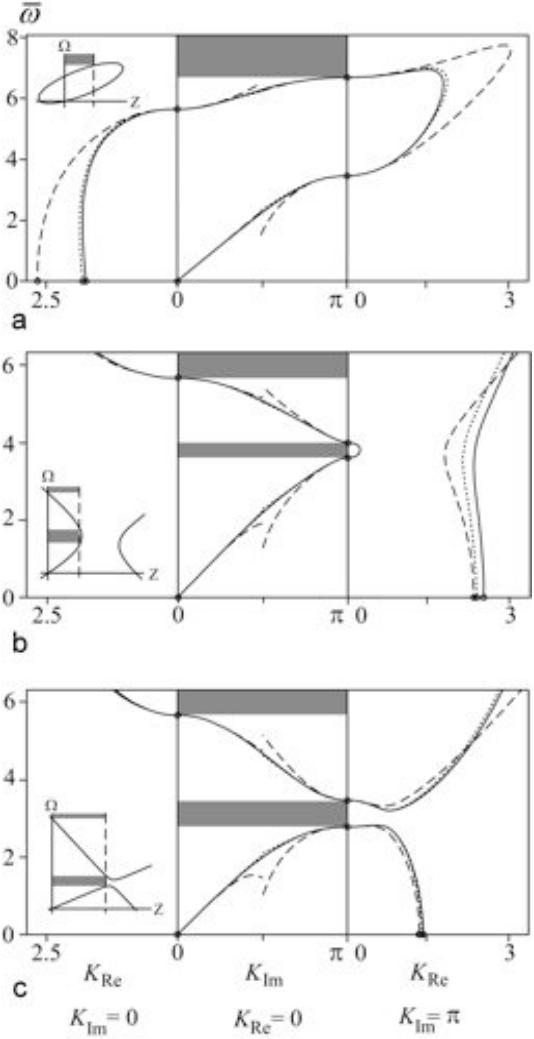

**Fig. 3.** Discrete system dispersion curves in $0\Omega Z$ axes (Vasiliev et al., 2010a) and in the planes $K_{\text{Im}} = 0$, $K_{\text{Re}} = 0$, and $K_{\text{Im}} = \pi$ (solid lines). Dispersion curves of 2FMm and 2FGMm are shown by dashed and dotted lines, respectively. Stop bands for the interfaces are shaded. 2FMm and 2FGMm give the same stop bands as the discrete model. Circles on the axis $K_{\text{Re}}$ in the plane $K_{\text{Im}} = 0$ (or $K_{\text{Im}} = \pi$) define the degree of localization for long- (or short-) wavelength boundary effects obtained for different models.

Let us note here that the error in the short-wavelength area may lead to instability of GMm. The error of this kind does not make the model useless because it can still be applied for analysis in the long-wavelength area. The instability problem has been discussed in several articles (see, for example, Askes and Aifantis (2011) and references therein) and we will not consider it here.

*4.3. Analysis of multi-field models*

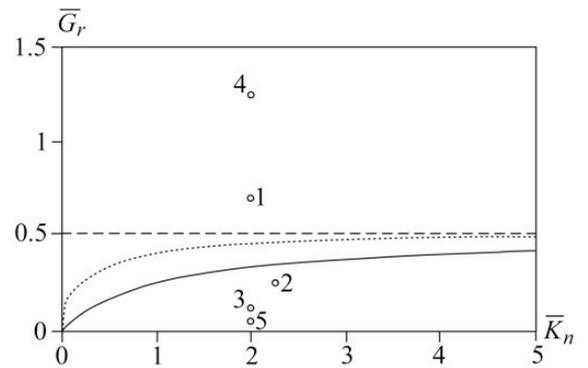

**Fig. 4.** The plane of dimensionless elastic parameters. The solid line divides system parameters into regions with exponential and short-wavelength boundary effects. Dispersion curves for the parameters marked by small circles "1","2", and "3" are shown in Fig. 3a, 3b, and 3c, respectively. Boundary effects in the interface with parameters "4","5" are shown in Fig. 7a and 7b.

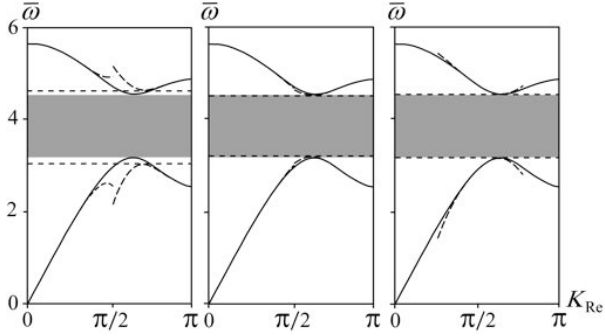

**Fig. 5.** Approximation of the harmonic wave dispersion curves of the discrete system (solid lines) by the dispersion curves of (a) 2FMm, (b) 2FGMm, and (c) 4FMm (dashed lines). Stop band for the interface defined by the discrete model is shaded. Stop bands defined by the multi-field models are shown by the horizontal lines.

$K_{\text{Im}} = \pi$ for the discrete model, 2FMm and 2FGMm are depicted by solid, dashed, and dotted lines for dimensionless parameters $\bar{K}_n = 2$, $\bar{G}_r = 0.7$; $\bar{K}_n = 2.25$, $\bar{G}_r = 0.25$; and $\bar{K}_n = 2$, $\bar{G}_r = 0.12$ in Figs. 3a, 3b, and 3c, respectively. These parameters are marked by small circles "1", "2", and "3" in Fig. 4. Figures 2 and 3a are drawn for the same parameters which makes it possible to compare the results for the single-field and two-field models. Dispersion curves of the single-field and two-field models coincide in the plane $K_{\text{Im}} = 0$ and in the half-interval $0 \leq K_{\text{Im}} < \pi/2$ of the plane $K_{\text{Re}} = 0$. They provide a good approximation for the dispersion curves of the discrete system at the long-wavelength region near the point $(K_{\text{Im}}, K_{\text{Re}}) = (0, 0)$. The two-field models improve the single-field models in the short-wavelength region, $K_{\text{Im}} \approx \pi$. The dispersion curves of the two-field models approximate the dispersion curves of the discrete system in the plane $K_{\text{Im}} = \pi$ and in the half-interval $\pi/2 < K_{\text{Im}} \leq \pi$ of the plane $K_{\text{Re}} = 0$ at the point $(K_{\text{Im}}, K_{\text{Re}}) = (\pi, 0)$.

Comparison of the multi-field models for harmonic waves is presented in Fig. 5. Here we consider parameters $\bar{K}_n = 5$, $\bar{G}_r = 0.1$. The dispersion curves of the discrete model are shown by the solid lines. The dispersion curves of 2FMm, 2FGMm, and 4FMm are shown by the dashed lines in Figs. 5a, 5b, and 5c, respectively. 2FGMm gives a better approximation than 2FMm at the points $K_{\text{Im}} = 0$ and $K_{\text{Im}} = \pi$. Both two-field models give maximal error at the point $K_{\text{Im}} = \pi/2$. The dispersion curves of 4FMm coincide with the dispersion curves of 2FMm on the half-intervals $0 \leq K_{\text{Im}} < \pi/4$ and $3\pi/4 < K_{\text{Im}} \leq \pi$. 4FMm gives the exact value of frequency at the point $K_{\text{Im}} = \pi/2$ and the second order approximation of dispersion curves of discrete system near this point.

### 4.4. Filtering property of structural interface

One of the most attractive features of periodic structures is their capability of attenuating elastic waves over certain frequency bands (Ruzzene and Scarpa, 2003, 2005). These bands are known as stop bands or band gaps. Filtering property of structural interfaces was studied by Bigoni and Movchan (2002) and by Brun et al. (2010). It is interesting to compare different continuum models in describing this property of the discrete system. Analysis of the dispersion relations of harmonic waves made in Subsections 4.1, 4.2, and 4.3 is important for finding the stop band edges. Localized solutions are interesting for the analysis of stop bands because they define the type and characteristics of spatial decay of waves.

The stop band edges are defined by (14) and (15) in the case when minima/maxima of the harmonic wave dispersion curves do not belong to the segment $0 \leq K_{\text{Im}} \leq \pi$. Dispersion curves of Mm and the discrete model coincide at the points (14). Mm and GMm both give considerable error at the points (15) in the short-wavelength region (Fig. 2). It was already noted that the dispersion curves of the discrete and two-field models coincide at (14) as well as at (15). We conclude that the two-field models can be used to find stop band edges of the discrete system. For the sake of illustration the stop bands are shaded in Fig. 3. The values (14) and (15) are marked by the small circles on the lines $K_{\text{Im}} = 0$ and $K_{\text{Im}} = \pi$. The stop bands of the discrete model and 2FMm coincide. The dispersion curves of the discrete and the two-field models in the case when minima/maxima belong to the segment $0 \leq K_{\text{Im}} \leq \pi$ are shown in Fig. 5a. The stop band defined by 2FMm is shown by the horizontal lines. 2FMm gives considerable error in the estimation of the stop band edges in this case.

The case when minima/maxima of dispersion curves belong to the segment $0 \leq K_{\text{Im}} \leq \pi$ requires additional analysis for the discrete system. Substituting $K = iK_{\text{Re}}$ into equation (13) and introducing new variables $\Omega = \omega^2$, $Z = 4\sin^2(K_{\text{Re}}/2)$ one can rewrite the dispersion relation $\omega = \omega(K)$ in the form $\Omega = \Omega(Z)$ (Vasiliev et al., 2010a). This form is useful for the analysis of the dispersion curves of the harmonic and localized dynamical solutions. The dispersion curves of harmonic waves $\omega = \omega(K_{\text{Im}})$, $0 \leq K_{\text{Im}} \leq \pi$, correspond to the branches of curves $\Omega = \Omega(Z)$ defined in $0 \leq Z \leq 4$. It was shown (Vasiliev et al., 2010a) that the curve $\Omega(Z)$ is ellipse, parabola, or hyperbola. Minimum/maximum of these curves belongs to the segment $0 \leq Z \leq 4$ if

$$\left.\frac{d\Omega}{dZ}\right|_{Z=0} \left.\frac{d\Omega}{dZ}\right|_{Z=4} < 0.$$

By using this relation one can find region of elastic parameters when minima/maxima of the dispersion curves belong to the segment $0 \le Z \le 4$. The acoustic branch $\Omega = \Omega(Z)$ has a maximum within the segment $0 \le Z \le 4$ for systems with elastic parameters in the region shown by the lines inclined by 45 degrees. The minimum/maximum of the optical branch belongs to the segment $0 \le Z \le 4$ for systems with parameters in the region shown by the lines inclined by 135 degrees. Curves are ellipses for systems with elastic parameters in the region between dotted lines and they are hyperbolas outside this region. Examples of typical curves for different regions are shown in Fig. 6 for points marked by circles "1" to "4" in the plane of elastic parameters.

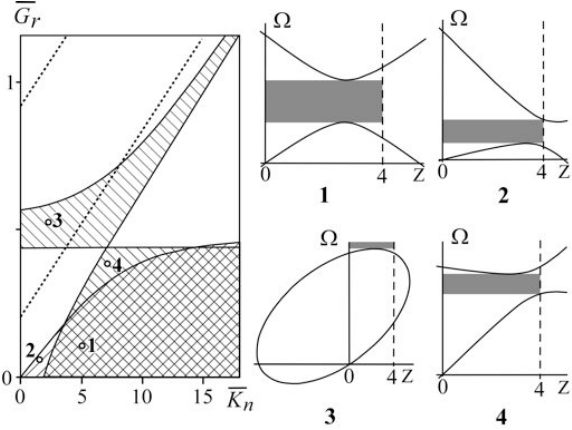

**Fig. 6.** The plane of dimensionless elastic parameters. Shaded are the regions of model parameters where minima/maxima of the discrete system dispersion curves belong to the interval $0 < K_{\text{Im}} < \pi$. The curves are ellipses for the systems with parameters between the dotted lines and they are hyperbolas outside this region. Examples of typical $\Omega(Z)$ curves for these regions with shaded stop bands.

It was noted that 2FMm may give a noticeable error in describing dispersion curves of the discrete system near the point $K_{\text{Re}} = \pi/2$ and, as a consequence, in finding minima/maxima of the dispersion curves and the stop band edges in the case when the minima/maxima belong to the segment $0 \le K_{\text{Im}} \le \pi$. On the other hand, the two-field model gives exact values for $\omega(K)$ and derivatives $d\omega(K)/dK$ of the discrete system dispersion curves at the boundaries $K_{\text{Im}} = 0$ and $K_{\text{Im}} = \pi$. By using the relation

$$\left.\frac{d\Omega}{dZ}\right|_{Z=4\sin^2(K_{\text{Im}}/2)} = \frac{\omega(K)}{\sin K}\left.\frac{d\omega(K)}{dK}\right|_{K=K_{\text{Im}}}$$

one can obtain the values $d\Omega/dZ|_{Z=0}$ and $d\Omega/dZ|_{Z=4}$ that enter the inequality (19). We come to an interesting result: although the two-field model can be inappropriate for evaluation of minima/maxima of the dispersion curves of discrete system, it can be used to determine if the minima/maxima belong to the segment $0 \le K_{\text{Im}} \le \pi$ and to find exactly the hatched regions in Fig. 6. 2GMm or 4FMm can describe dispersion curves of the discrete system more accurately.

Comparison of the multi-field models in the case when maxima/minima of the dispersion curves belong to the segment $0 \le K_{\text{Im}} \le \pi$ is given in Fig. 5, where the stop band is shaded. Stop band edges defined by the multi-field models are shown by horizontal lines. One can see that the stop band defined by 2FMm model is wider than the stop band of the discrete system (Fig. 5a). 2FGMm (Fig. 5b) and 4FMm (Fig. 5c) provide an appropriate level of accuracy.

## 5. Static solutions. Long- and short-wavelength boundary effects

Static solutions are defined by the values of $K$ when the dispersion curves (13) intersect the plane $\omega = 0$, i.e.,

$$\det\begin{bmatrix} a_1 & -a_2 \\ a_2 & a_3 \end{bmatrix} = 0. \tag{20}$$

Analysis of static solutions for the Cosserat lattice has been done in Vasiliev et al. (2010a). Static solutions consist of linear and spatially localized parts.

The characteristic equation (20) has the double root $K = 0$ that defines the linear part of static solution. The type of two other solutions to the characteristic equation is defined by the parameter $\gamma = (\delta + 1)/(\delta - 1)$, where $\delta = 2(K_n + K_s)G_r / K_n K_s d^2$. The solutions are real $K_{\text{Re}}^\pm = \ln\left[\gamma \pm \sqrt{\gamma^2 - 1}\right]$ for $\delta > 1$ and complex $K^\pm = \ln\left[|\gamma| \mp \sqrt{\gamma^2 - 1}\right] + i\pi$ for $0 < \delta < 1$. In the former case, we have exponentially localized deformations and the general solution has the form

$$v_m = C_1 + 2mdC_2 + e^{K_{\text{Re}} m}C_3 + e^{-K_{\text{Re}} m}C_4,$$
$$\varphi_m = C_2 + \alpha k\left[e^{K_{\text{Re}} m}C_3 - e^{-K_{\text{Re}} m}C_4\right], \tag{21}$$

where

$$k = (K_n + K_s)/dK_s, \quad \alpha = (\cosh K_{\text{Re}} - 1)/\sinh K_{\text{Re}}.$$

In the latter case, the short-wavelength localized deformations of the following form take place

$$v_m = C_1 + 2mdC_2 + (-1)^m e^{K_{\text{Re}} m}C_3 + (-1)^m e^{-K_{\text{Re}} m}C_4,$$
$$\varphi_m = C_2 + \alpha k\left[(-1)^m e^{K_{\text{Re}} m}C_3 - (-1)^m e^{-K_{\text{Re}} m}C_4\right], \tag{22}$$

where $\alpha = (\cosh K_{\text{Re}} + 1)/\sinh K_{\text{Re}}$.

The parameter

$$K_{\text{Re}} = \ln\left[|\gamma| + \sqrt{\gamma^2 - 1}\right] > 0 \tag{23}$$

defines the degree of spatial localization of the boundary effect in the solutions (21), (22). Smaller value of the parameter means weaker localization.

Characteristic polynomial (20) of Mm with $c_m$ defined by (17) gives four roots. The double root $K = 0$ gives the linear part of the solution. Two other roots,

$$K_{\text{Re}} = \pm\sqrt{4K_n K_s d^2 / (2G_r - K_s d^2)(K_n + K_s)}, \quad (24)$$

are real for the interfaces with parameters $G_r / K_s d^2 \geq 1/2$ ($\overline{G}_r \geq 1/2$ in Fig. 4). They lead to spatially localized solutions. The general static solution to Mm equations (3) has the form

$$V^{[1]}(\xi) = C_1 + 2\xi C_2 + e^{K_{\text{Re}}\xi/d} C_3 + e^{-K_{\text{Re}}\xi/d} C_4,$$
$$\Phi^{[1]}(\xi) = C_2 + \alpha k \left[ e^{K_{\text{Re}}\xi/d} C_3 - e^{-K_{\text{Re}}\xi/d} C_4 \right], \quad (25)$$

where $\alpha = (K_{\text{Re}}^2/2)/K_{\text{Re}}$.

Solution (25) has the form of the discrete solution (21). Parameter $K_{\text{Re}}$ in both solutions is defined by the equation (20) with components $c_m$ for Mm, Eq. (3), and with components $a_m$ for the discrete model, Eq. (2). Since components $c_m$ are nothing but Taylor series expansions of the components $a_m$, the parameter $K_{\text{Re}}$ defined by Mm is close to that of the discrete system in the case $K_{\text{Re}} \approx 0$, i.e., for weak localization. For the numerator and denominator of the parameter $\alpha$ in solutions (21) and (25), we have

$$\cosh K_{\text{Re}} - 1 = K_{\text{Re}}^2 / 2 + o(K_{\text{Re}}^4),$$
$$\sinh K_{\text{Re}} = K_{\text{Re}} + o(K_{\text{Re}}^3).$$

Thus, solution (25) for Mm describes displacements in the discrete system with a higher accuracy for smaller values of $K_{\text{Re}}$, i.e., when the displacements vary slowly.

The degree of approximation for the components $a_m$ and for the numerator and denominator of the parameter $\alpha$ is of fourth order in the case of GMm, Eq. (7). Exponentially localized solutions obtained for the discrete model, for Mm and for GMm exist for elastic parameters in the regions above the solid, dashed, and dotted lines in Fig. 4. Thus, GMm is more accurate than Mm and it gives solutions in a wider range of the elastic parameters. However, GMm is more complicated and its use can be recommended only in case of strongly localized solutions.

Mm and GMm do not describe the short-wavelength localized solutions (22) that exist for structural interfaces with elastic parameters $0 \leq 2(K_n + K_s)G_r / K_n K_s d^2 < 1$ (below the solid line in Fig. 4).

2FMm includes equations (3) of Mm. These equations give the solution (21) for the fields $V^{[1]}(\xi)$, $\Phi^{[1]}(\xi)$. Equations (7) of 2FMm give $V^{[2]}(\xi) = 0$, $\Phi^{[2]}(\xi) = 0$ in this case. Taking into account the relations (6), we find that the pairs of approximating fields of 2FMm coincide with each other and with the fields defined by the single-field model

$$v^{[1]}(\xi) = v^{[2]}(\xi) = V^{[1]}(\xi),$$
$$\varphi^{[1]}(\xi) = \varphi^{[2]}(\xi) = \Phi^{[1]}(\xi).$$

Thus, 2FMm gives the accuracy of Mm in describing the exponential boundary effects.

Moreover, 2FMm can describe the short-wavelength boundary effects. One can derive linear part of solution by using equations (3) of 2FMm for $V^{[1]}(\xi)$, $\Phi^{[1]}(\xi)$. The short-wavelength static solutions can be derived by using equations (7). Localization parameter $K_{\text{Re}}$ is a real solution to the characteristic equation (20) with components $\tilde{c}_m$. Equation (7) gives the corresponding solutions $V^{[2]}(\xi)$, $\Phi^{[2]}(\xi)$. Taking into account the relations (6), we obtain the approximating fields $s = 1, 2$

$$v^{[s]}(\xi) = C_1 + 2\xi C_2 +$$
$$+ (-1)^s e^{K_{\text{Re}}\xi/d} C_3 + (-1)^s e^{-K_{\text{Re}}\xi/d} C_4,$$
$$\varphi^{[s]}(\xi) = C_2 +$$
$$+ \alpha k \left[ (-1)^s e^{K_{\text{Re}}\xi/d} C_3 - (-1)^s e^{-K_{\text{Re}}\xi/d} C_4 \right] \quad (26)$$

where $\alpha = (K_{\text{Re}}^2/2 + 2)/K_{\text{Re}}$.

This solution is similar to the discrete solution (22). One can show that 2FMm predicts the localization parameter $K_{\text{Re}}$ with a high accuracy in the case of weak localization, i.e., when $K_{\text{Re}} \approx 0$. Indeed, solutions (22) and (26) are close for small $K_{\text{Re}}$ because for the numerator and denominator of the parameter $\alpha$ one has

$$\cosh K_{\text{Re}} + 1 = K_{\text{Re}}^2 / 2 + 2 + o(K_{\text{Re}}^4),$$
$$\sinh K_{\text{Re}} = K_{\text{Re}} + o(K_{\text{Re}}^3).$$

2FGMm gives a better approximation.

The localization parameter $K_{\text{Re}}$ is one of the important characteristics of the considered structural interface. Figure 3 illustrates the accuracy of 2FMm and 2FGMm in estimation of this parameter. The values $K_{\text{Re}}$ are shown by the circles at the edges of the dispersion curves on the line $\omega = 0$. If the point of intersection belongs to the plane $K_{\text{Im}} = 0$ (Fig. 3a), then we have exponential localization. Short-wavelength localization takes place if this point belongs to the plane $K_{\text{Im}} = \pi$ (Figs. 3b, 3c). One can see that 2FGMm provides a better accuracy in determination of the localization parameter $K_{\text{Re}}$ than 2FMm.

Figures 7a and 7b illustrate the approximation of exponential and short-wavelength boundary effects by the two fields of 2FMm for the values of parameters $\overline{K}_n = 2$, $\overline{G}_r = 1.25$ and $\overline{K}_n = 2$, $\overline{G}_r = 0.05$, respectively (the points "4" and "5" in Fig. 4). We consider shear deformations of the structural interface consisting of 15 layers of particles applying the boundary conditions $v_{-7} = -v_*$, $\varphi_{-7} = 0$, $v_7 = v_*$, $\varphi_7 = 0$, where $v_* = 0.05d$. The values $\varphi_m$ in the layers $m = \overline{-7\ldots 7}$ are shown by small circles connected by the solid line to reveal the character of deformations. Approximating fields $\varphi^{[1]}(\xi)$, $\varphi^{[2]}(\xi)$ of 2FMm are shown by the dashed lines. As it can be seen, both slowly varying and short-wavelength deformations can be accurately described by the two slowly varying functions. These functions coincide in the case of slowly varying deformations (Fig. 7a) and 2FMm gives the accuracy of Mm. Solid line shows that single function should be fast varying in order to describe short-wavelength deformations (Fig. 7b). Mm and GMm cannot be used to describe short-wavelength deformations because only the lowest order terms in the Taylor series expansions are taken into account in these models. Influence of the high-order gradient terms is small for the long-wavelength deformations but it is not small in the case when wavelength is comparable with the size of a periodic cell, i.e. for rapidly varying deformations. In the two-field theory, short-wavelength deformations are accurately described by the two slowly varying field functions $\varphi^{[1]}(\xi)$, $\varphi^{[2]}(\xi)$ (Fig. 7b). Long-wavelength theory gives a good accuracy for such fields.

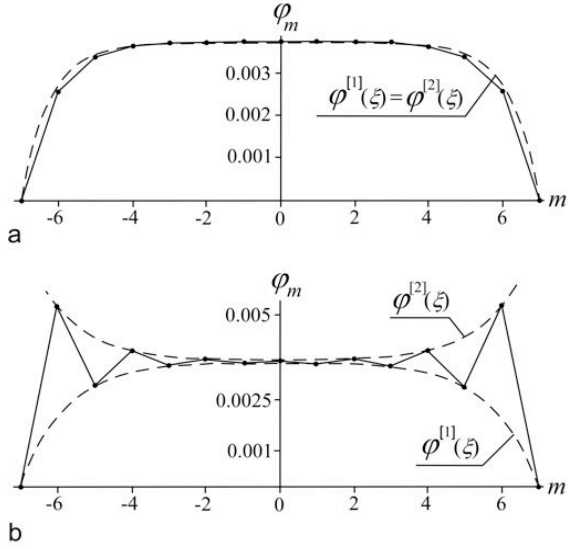

**Fig. 7.** The values of $\varphi_m$ for shear deformations in the interface with (a) exponential and (b) short-wavelength boundary effects (circles connected by the solid line). Approximations by the two slowly varying functions $\varphi^{[1]}(\xi)$ and $\varphi^{[2]}(\xi)$ of 2FMm (dashed lines).

## 6. Conclusion

The generalized continuum mechanics has as an aim development of conventional mechanics by enriching it by new ideas, conceptions, models, and new fields of application. It is naturally that one of the central directions of development is the study of multi-scale problems. Another direction is correct description of multi-physics problems for complex materials when the fields of different physical nature and their interactions are considered. Such problems take an increasing interest (see, e.g., Mariano, 2002; Borst, 2008; Sansour and Skatulla, 2009; Geers et. al, 2010; Maugin, 2010; Vasiliev et al., 2010b; Askes and Aifantis, 2011; Aifantis, 2011; Li and Ostoja-Starzewski, 2011; Charlotte and Truskinovsky, 2008, 2012).

In the present study we focus of the first of the above-mentioned problems and obtain several generalized continuum models and discuss their applicability to modeling some important effects in the structural interface modeled by the Cosserat lattice, each particle of which possesses translational and rotational degrees of freedom. Equations of Mm are derived by Taylor series expansion up to the second order terms of the discrete equations of motion written for the particles within single translational cell with respect to displacements and rotations. Construction of GMm relies on retaining the higher-order terms (up to the fourth order terms in our case) in the Taylor series expansions. We derive $n$-field micropolar model considering macrocell consisting of $n$ translational cells and using $n$ vector fields to describe the deformations of the lattice. Derivation of the $n$-field gradient micropolar model implies the combination of the two generalization approaches, namely, $n$ vector fields are used in order to describe deformations and the higher-order terms are taken into account in the Taylor series expansions of the discrete equations written for $n$ translational cells of the lattice.

Not only harmonic but also spatially localized solutions were derived and analyzed. The localized solutions can play important role for the interfaces of relatively small thickness. The dispersion curves of the derived continuum models are compared to that of the discrete model in order to estimate the accuracy of the continuum models. It was demonstrated that Mm can be applied to study the long-wavelength harmonic waves, as well as the static and dynamic spatially decaying solutions with weak localization. In addition to this, GMm describes the dispersion of waves and strongly localized solutions. However, the single-field models give considerable error for short-wavelength harmonic solutions and do not reproduce localized short-wavelength solutions. 2FMm and 2FGMm include equations of Mm and GMm and they have two more equations to capture the short-wavelength waves. 4FMm includes equations of 2FMm and it has four more equations to improve the description of the waves in the middle part of the first Brillouin zone. Thus, the multi-field models describe not only long-wavelength deformations but also short-wavelength effects where the single-field theories give considerable error and cannot be used.

As one of the possible applications of the multi-field theories we consider the filtering properties of structural solids finding the stop band edges. We show that the two-field models make it possible to describe not only long-wavelength but also short-wavelength boundary effects in the structural interface and evaluate the degree of its spatial localization.

Let us note some other problems where our results may be useful and name several directions for future studies.

Continuum models are effectively used to study nonlinear phenomena in discrete systems. For example, an application of continuum models based on Pade approximations for the study of discrete kinks was considered by Kevrekidis et al. (2002), Andrianov et al. (2012). Multi-field models were obtained in Dmitriev (1997) in order to find moving kinks connecting short-wavelength discrete *n*-periodic structures. Careful description of stop bands and short wavelength solutions is important in linear analysis of spatially localized short-wavelength excitations having frequencies within the stop bands. Such excitations are known as discrete breathers and rotobreathers (Flach and Willis, 1998; Kevrekidis et al., 2004). Study of the discrete breathers and rotobreathers in Cosserat lattices is an interesting problem. The fact that the multi-field theories describe both long and short wavelength excitations makes it possible to apply them to the analysis of interaction of excitations at macro- and micro-structural levels in frame of continuum mechanics.

In the present article we start from a full Cosserat model and consider complex generalized continuum models in order to describe static and dynamic behavior of Cosserat lattice with increasing accuracy. However, it is interesting to consider the models simpler than the micropolar model. For example, for lattices with parameters $2G_r = K_s d^2$ derivatives of the second order for rotations in Eqs. (3), (4), and (7) are absent and we have the reduced Cosserat model (Grekova et al., 2009) and its generalizations. In theoretical studies, the elimination of rotational degrees of freedom may help to better understand the theories where only translation degrees of freedom are considered and the nature of spatially and temporary nonlocal (gradient) models.

In the present article we paid main attention to the localized static and dynamic solutions. We considered harmonic solutions in order to find the stop band edges and to describe filtering properties that are important for interfaces, its analysis and applications. Certainly, the results of our analysis are applicable for infinite bodies. However, in the latter case most attention is typically paid to harmonic solutions, group velocity and other macro characteristics. The reflection/transmission at the interface-material boundaries, the surface waves, and the role of the short-wavelength boundary effects in fracture are the interconnected problems for future studies.

**Acknowledgments**

A.A.V. acknowledges the partial support of the Ministry of Education and Science of Russian Federation (State contract No. 07.524.11.4019 for development). S.V.D. appreciates financial support from the Russian Foundation for Basic Research, grants 11-08-97057-p_povolzhie_a and 10-08-90012-Bel_a.